# Charting the Future of AI-supported Science Education: A Human-Centered Vision

Xiaoming Zhai[1], Kent Crippen[2]

[1]*AI4STEM Education Center, University of Georgia*

[2]*School of Teaching and Learning, University of Florida*

This concluding chapter explores how artificial intelligence (AI) is reshaping the purposes, practices, and outcomes of science education, and proposes a human-centered framework for its responsible integration. Drawing on insights from international collaborations and the Advancing AI in Science Education (AASE) committee, the chapter synthesizes developments across five dimensions: educational goals, instructional procedures, learning materials, assessment, and outcomes. We argue that AI offers transformative potential to enrich inquiry, personalize learning, and support teacher practice, but only when guided by Responsible and Ethical Principles (REP). The REP framework—emphasizing fairness, transparency, privacy, accountability, and respect for human values—anchors our vision for AI-supported science education. Key discussions include the redefinition of scientific literacy to encompass AI literacy, the evolving roles of teachers and learners in AI-supported classrooms, and the design of adaptive learning materials and assessments that preserve authenticity and integrity. We highlight both opportunities and risks, stressing the need for critical engagement with AI to avoid reinforcing inequities or undermining human agency. Ultimately, this chapter advances a vision in which science education prepares

---

Zhai, X. & Crippen, K. (2026). Charting the Future of AI-supported Science Education: A Human-Centered Vision. In X. Zhai & K. Crippen (Eds.). Advancing AI in Science Education: Envisioning Responsible and Ethical Practice. Springer.

learners to act as ethical investigators and responsible citizens, ensuring that AI innovation aligns with human dignity, equity, and the broader goals of scientific literacy.

*Keywords*: Artificial Intelligence (AI); Advancing AI in Science Education (AASE); Responsible and Ethical Principles; Human-Centered

# 1. Introduction

We are living in an era of rapid transformation, in which science education must adapt to the evolving demands of a workforce that requires discipline-based competencies in artificial intelligence (AI). Unlike earlier waves of technological change that often built incrementally upon established practices, the present shift is more radical and disruptive, demanding not only new skills but also integrated competence that can thrive with the technology itself. Moreover, the pervasive nature of AI in society implies that students encounter these technologies not only in professional contexts but also in their civic and personal lives, making it imperative for education systems to nurture critical awareness, ethical reasoning, and the ability to question and evaluate AI's role in shaping knowledge and decision-making. To meet this change, our educational system must make substantial adjustment at various levels, ranging from the National AI Strategy to reimagined paradigms of teaching and learning (Shi, 2025). This transition is neither trivial nor paradigmatic, yet it is fundamentally reshaping how we conceive of teaching, learning, and assessment in science education. It challenges educators, researchers, and policymakers to rethink the very purposes and methods of science education in light of these shifts, underscoring the urgency for science education to prepare learners to engage with AI technologies in responsible, ethical, and human-centered ways.

The field of science education is navigating this transformation along three interrelated trajectories, each of which reflects both historical continuities and novel possibilities. First, new scholarly communities are emerging to cultivate research and practice at the intersection of AI and science education. A notable example is the Research on AI-involved Science Education (RAISE) special interest group within the NARST organization. Founded in 2022, RAISE has provided a professional home for collaboration and intellectual exchange ( Zhai, 2024). Beyond functioning as a network, RAISE has initiated workshops, conference sessions, and mentoring opportunities that have allowed researchers from diverse disciplinary backgrounds to coalesce around shared questions, methods, and values. In this way, the community is actively shaping the identity of a new research subfield rather than merely reacting to technological trends (Lee et al., 2025). Over time, such communities not only produce knowledge but also establish norms, standards, and agendas that shape the trajectory of scholarship and practice for years to come.

Second, a growing body of research is generating foundational knowledge on the potential applications of AI in science education. This scholarly activity extends beyond documenting novel uses of AI technologies and increasingly seeks to theorize how AI reshapes learning processes, instructional design, and teacher professional development. Examples of this activity include the 2021 special issue, titled *Applying Machine Learning in Science Assessment* in the *Journal of Science Education and Technology*, which set forth a new vision for the potential of AI, in particular to advance next-generation science assessment. The volume *Uses of Artificial Intelligence in STEM Education* (Zhai & Krajcik, 2025b), which emerged from the *International Conference on AI-based Assessment in STEM Education* held in Athens, Georgia, in 2022, sets a milestone that exemplifies more than 60 scholar's cutting-edge work in AI-supported science education. These scholarly efforts not only compiled empirical evidence but also articulated visions for integrating AI in sustainable and scalable ways. Together, these contributions mark a transition from early exploratory studies toward more systematic and programmatic lines of inquiry. In addition, newer projects are beginning to develop design principles, empirical frameworks, and methodological innovations that will guide subsequent generations of researchers and practitioners. By doing so, this body of work is gradually moving from curiosity-driven exploration toward the establishment of a coherent, evidence-informed field with its own theoretical foundations.

Next, the community has begun to confront the challenges and ethical implications of AI in science education. While enthusiasm about AI's potential is evident, critical scholarship has emphasized that adoption without deep reflective thought and consideration risks reinforcing inequities, narrowing pedagogical aims, or diminishing the central role of human agency in science learning. Along this line, confusion between AI and the use of AI, misunderstanding of AI and its mechanisms and outcomes have been constantly drawing concerns (Bewersdorff et al., 2023; Zhai & Krajcik, 2025a), not only among practitioners, but also within scholarly work. For example, recent debates published in the *Journal of Research in Science Teaching* (Krist & Kubsch, 2023; Li et al., 2023; Zhai & Nehm, 2023) exemplify several alternative understandings of AI in science education, yet collectively highlight both the promise and the perils of AI. These papers call for greater attention to issues of responsibility, equity, and ethics when using AI in science assessment. These discussions clarify that AI technology is not a neutral tool, but an embedded element in a socio-technical system that includes cultural values, power dynamics, and policy environments. These debates also point to the need for more cross-disciplinary dialogue with ethicists, policymakers, and practitioners, ensuring that the integration of AI into science education is more than driven by technology, but a process that must be informed by broader societal concerns and long-term humanistic goals.

In response to this broadening conversation, the Advancing AI in Science Education (AASE) committee was established under the leadership of the co-editors, Xiaoming Zhai and Kent Crippen. Funded by National Science Foundation, the committee was charged with two primary goals: to identify the transformations AI brings to science education and to articulate principles for responsible and ethical practice of AI in science education. These goals were intentionally ambitious, recognizing that addressing them would require not only the synthesis of current research but also the creation of new opportunities for dialogue across disciplinary and institutional boundaries. This work unfolded through multiple virtual meetings, which allowed for the participation of international scholars, and culminated in a two-day in-person conference held in Athens, Georgia, in February 2025. That conference served as both a capstone and a catalyst: it provided an opportunity to refine the committee's shared understandings, generate new ideas, and engage in spirited debate about the implications of AI for the future of science education. The deliberations and collaborations of the committee produced the collective insights that structure this volume, as well as a sense of community that extends beyond the book itself. The book includes ten chapters authored by the committee members, in addition to introductory and concluding chapters, that together address the committee's charge. In doing so, it aspires not only to document the state of the field but also to serve as a touchstone for future scholarship and practice, a roadmap for teacher educators and policymakers, and an invitation to practitioners who are exploring the promises and pitfalls of AI in their own contexts.

Our approach to this endeavor began with examining the shifts AI has precipitated in contemporary scientific research and considering how these shifts inform science education (see Chapter 2). Building on this foundation, we proposed a framework for responsible and ethical principles (REP) of practicing AI-supported science education (see Chapter 3). This framework composites principles covering eight aspects: *Fairness, Privacy & Data Security, Human Oversight & Autonomy, Beneficial & Meaningful Use, Robustness & Scientific Integrity, Respect for Human Values, Transparency & Explainability,* and *Accountability*. The framework was not conceived in abstraction, but rather, grounded in careful analysis of ongoing case studies and collaborative reflection on best practices. Guided by this framework, the book is organized into five sections, each examining how AI is reshaping core dimensions of science education—*educational goals, pedagogical processes, learning materials, assessment practices,* and *educational outcomes*—and articulate ethical and responsible practice of AI guided by the REP framework. Each section is intended to provide conceptual clarity and practical guidance, demonstrating ways in which AI can be used to expand opportunities for learners while safeguarding human dignity and agency. In what follows, we present the key insights that emerged from the committee's work, charting a vision for the future of AI in science education that is both innovative and human-centered. These insights provide a foundation upon which subsequent scholarship, classroom innovations, and policy initiatives may build and evolve in the years to come.

## 2. Key Insights from the Book's Sections

### 2.1. Educational Goals

The formulation of educational goals must align with the evolving demands of future citizens and the workforce. In light of the transformative influence of AI across all sectors of society, it is evident that the competencies required for individuals to thrive are undergoing profound change. Consequently, educational objectives—particularly those embedded within science education—must be redefined to address these emerging needs. Within this context, two chapters are dedicated to examining the necessary transformations (see Chapters 4 and 5). The combined insights from these chapters highlight the following:

*Evolving Purpose of Science Education*

> Science literacy has expanded beyond content mastery to include applying knowledge in everyday life and exercising agency in addressing societal issues. In the AI era, these goals converge into preparing students as lifelong investigators and responsible citizens.

*AI as a Transformative Force in Science and Education*

> Artificial intelligence is reshaping how science is practiced—accelerating inquiry, modeling, experimentation, and data analysis—and requires K–12 education to integrate AI as an authentic component of scientific practice.

*Science Literacy Plus as a New Educational Goal*

> A future-ready framework combines science literacy and AI literacy. It emphasizes human-centered mindsets, ethical engagement, technical proficiency, and system design skills, progressing from understanding to applying to creating.

*Inquiry-Centered, AI-Supported Learning*

> Science classrooms should move from rote learning toward investigation-centered environments where AI serves as a partner in problem-solving, enabling authentic inquiry cycles that emphasize evidence, iteration, creativity, and reflection.

*Developing Reflective and Ethical Competencies*

> Students must cultivate metacognitive, critical, and ethical skills to engage responsibly with AI, addressing risks such as bias, misinformation, inequity, and privacy concerns while promoting transparency, fairness, inclusivity, and environmental responsibility.

*Personalized and Inclusive Learning with AI*

> AI tools, such as natural language processing and adaptive guidance, can personalize feedback, make student thinking visible, and build on diverse learner ideas, supporting equitable and meaningful participation in science learning.

*Collaboration and Partnerships for Sustainable Integration*

> Equitable and ethical AI integration in education requires partnerships among teachers, researchers, communities, and policymakers to co-develop open-source curricula, resources, and professional learning opportunities.

*Preparing Autonomous and Resourceful Learners*

> By embracing AI-supported inquiry and Science Literacy Plus, K–12 education can prepare students to be self-directed, collaborative, and ethically responsible learners capable of addressing urgent scientific and societal challenges.

### 2.2. Educational Procedures

As educational goals evolve, there is an urgent need to reconfigure instructional practices in science classrooms and to prepare teachers with the knowledge and skills necessary to navigate AI-enhanced learning environments (X. Zhai, 2024). Modern science education is increasingly designed to foster scientific practices and collaborative problem-solving. With the integration of AI, these processes are further amplified through innovative instructional decision-making. AI-supported systems can provide immediate feedback, simulate experimental conditions, and personalize learning pathways, enabling students to generate and test hypotheses, interpret results, and progressively refine their scientific understanding. In this context, teachers' roles are being redefined (Shi & Choi, 2024), and instructional practices are shifting significantly from those of traditional classrooms. Moreover, AI holds the potential to enhance the overall effectiveness of teaching and learning. The two chapters in this section address these transformations, with particular attention to the responsible and ethical use of AI in education.

*Individualized and Authentic Student Learning*

> AI enables teachers to tailor instruction through intelligent tutoring systems, exploratory environments, and adaptive curricula, while generative AI can scaffold authentic, self-directed, and agentic learning experiences.

*Shifting Classroom Dynamics and Roles*

> The integration of AI requires a move from teacher-centered orchestration to teacher–student–AI co-orchestration, with teachers acting as facilitators, organizers, and critical evaluators of AI contributions.

*Balancing Autonomy and Guidance*

> AI can scaffold inquiry and reasoning but also risks undermining independent sense-making if adopted uncritically. Students must develop epistemic autonomy, metacognition, and reflective skills to engage responsibly with AI outputs.

*Ethical and Equitable Integration*

Responsible use of AI requires instructional strategies, collaboration frameworks, and policies that ensure fairness, transparency, inclusivity, privacy protection, and human oversight in both teaching and learning.

*Building Shared AI Literacy*

Both teachers and students must cultivate AI literacy—not only to use AI effectively but also to critically evaluate outputs, recognize limitations, and address ethical challenges in algorithmic mediation.

*Amplifying Human Aspects of Education*

When implemented thoughtfully, AI should enhance—not replace—the human dimensions of science teaching and learning, such as curiosity, inquiry, creativity, and meaningful teacher–student connections.

*AI as a Tool for Teacher Support and Efficiency*

AI can streamline lesson planning, instruction, assessment, and reflection while providing timely insights into student knowledge and struggles, allowing teachers to focus on responsive and targeted instruction.

*Enhancing Teacher Professional Learning*

AI offers practice-based opportunities, personalized feedback, and simulations that strengthen teachers' competencies. Professional development must build AI literacy, model effective use of prompts, and emphasize teacher judgment as central to decision-making.

*2.3. Learning Materials*

If educational goals and procedures are being reshaped by AI, the demand for transformed learning resources in science education becomes evident. This transformation is both profound and multifaceted, emphasizing interactive, adaptive, and multimodal formats. Traditional static textbooks are increasingly being replaced by virtual laboratories, AI-supported simulations, and augmented reality environments that enable learners to engage directly with scientific phenomena. To ensure that science education remains contemporary, content of the learning material must extend beyond conventional methods and reflect the authentic practice employed by scientists. Moreover, AI plays a growing role in supporting educators and researchers in the creation of relevant instructional materials, reducing the demands of time-intensive tasks and allowing curriculum designers to focus more fully on higher-order pedagogical innovation and design. Chapter 8 contributes to the field in several important ways:

*Embedding Authentic Scientific Practice*

> AI transforms science learning materials by integrating practices such as data analysis, modeling, simulation, and hypothesis testing, enabling students to engage with science as it is practiced today.

*Personalization and Adaptivity*

> AI-supported materials can dynamically respond to learners' needs, prior knowledge, and progress, fostering inclusivity, deeper engagement, and equitable learning opportunities.

*Interactive and Inquiry-Rich Experiences*

>AI-supported simulations and tools allow students to become active investigators and even co-creators of models and experiments, providing contextualized, authentic, and inquiry-rich learning experiences.

*Multimodal and Dynamic Content*

>Generative AI enables the creation of multimodal resources (text, visuals, audio, video) that make abstract concepts more tangible, but their educational use requires careful validation for scientific accuracy and pedagogical coherence.

*Accessibility and Cultural Responsiveness*

>AI enhances accessibility for multilingual, neurodiverse, and differently abled learners through translation, summarization, and multimodal interaction, while supporting culturally responsive and inclusive learning design.

*Co-Authoring and Rapid Content Development*

>Generative AI can serve as a collaborative partner for educators in creating context-sensitive, culturally relevant, and adaptable instructional resources, provided that human oversight ensures quality, integrity, and alignment with learning goals.

*Ethical and Responsible Integration*

>The use of AI in learning materials must prioritize transparency, fairness, privacy, inclusivity, and human oversight, ensuring that teacher judgment and student agency remain central.

*Capacity Building for Educators and Learners*

> Professional development is essential for teachers to evaluate, adapt, and design AI-supported materials, while students must develop AI literacy to critically interpret, question, and contextualize AI outputs.

*Reimagining Science Learning Materials*

> When implemented responsibly, AI can act as a catalyst to make learning materials more authentic, inclusive, and responsive, advancing the broader goals of scientific literacy, critical thinking, and ethical responsibility.

## 2.4. Assessment

AI is also transforming approaches to assessment in science education, shifting emphasis toward more formative, performance-based, and authentic measures of learning. Automated scoring systems can now evaluate students' written explanations, arguments, and drawn models, enhancing efficiency while expanding the scope of constructs that can be assessed. Through AI-supported learning analytics, educators can monitor how students plan investigations, interpret evidence, and construct scientific arguments, providing insight into both the processes and outcomes of science learning. Interactive dashboards supply real-time feedback on student learning progressions of big ideas in science, enabling timely instructional responses. Furthermore, AI-supported longitudinal assessments compile detailed learner profiles, allowing educators to trace the progression of scientific reasoning over extended periods. In this way, assessment becomes a dynamic and integral element of the learning process rather than a concluding event. Two chapters dedicated to this section articulates both the potential and the challenges of AI-supported science assessment.

*Expanding Assessment Constructs for the AI Era*

Science assessment must evolve to capture competencies relevant to AI-supported learning, including collaboration with AI, critical evaluation of AI outputs, ethical reasoning, and higher-order problem-solving.

*Designing AI-Resilient and Authentic Assessments*

Traditional recall-based tasks are increasingly vulnerable to AI-supported shortcuts. Assessments should emphasize performance-based, inquiry-driven, and higher-order thinking tasks that preserve authentic student engagement and align with standards.

*Balancing Efficiency with Integrity*

While AI can streamline scoring and feedback, overreliance by students (outsourcing responses) or teachers (outsourcing item design and scoring) risks undermining meaningful learning, assessment literacy, and instructional judgment.

*Ensuring Validity, Fairness, and Oversight*

AI-supported scoring requires transparent rubrics, continuous validation, and human-in-the-loop oversight to safeguard interpretability and fairness across diverse populations. Frequent recalibration is essential to maintain educational value.

*Mitigating Bias through Technical and Societal Strategies*

Both societal and technical biases must be addressed. This includes auditing systems for equity, using balanced and diverse training datasets, designing analytic rubrics, and adopting open-source or transparent AI tools to improve accountability.

*Accounting for Locality and Context*

> Assessment practices must consider standards, language, culture, region, and accessibility. Strategies such as retrieval-augmented generation for curriculum alignment, multilingual model training, and culturally responsive design are necessary for inclusivity.

*Transforming Classroom Assessment into Formative Practice*

> AI can enable ongoing, formative assessment by providing real-time feedback, multimodal data analysis, and learner-driven engagement, fostering more personalized and responsive instruction.

*Fostering AI Literacy and Ethical Engagement*

> Both teachers and students need AI literacy to understand its affordances, limitations, and ethical implications. Human judgment, reflection, and responsibility must remain central to all AI-supported assessment practices.

*Positioning AI as a Catalyst, not a Substitute*

> AI should be leveraged to enrich assessment by deepening inquiry, supporting fairness, and amplifying human expertise—rather than as a shortcut for efficiency or replacement of authentic teacher and student roles.

### 2.5. Educational Outcomes

In the end, shifts in goals, procedures, learning resources, and assessment practices reshape what counts as achievement in science education. Success is no longer measured solely by memorizing facts or completing lab tasks; instead, it highlights originality in designing experiments, teamwork across disciplines, persistence in tackling challenges, and thoughtful consideration of the social and ethical implications of science and technology, including AI. Graduates of AI-supported science education are imagined as flexible, lifelong learners, prepared to confront both scientific and societal issues responsibly—for example, by advancing renewable energy or addressing climate change. This reimagining ensures that science education goes beyond delivering subject knowledge, fostering the broader capabilities needed to drive innovation in an AI-infused world.

*Preparing Students for an AI-Infused World*

> As AI transforms science, work, and daily life, K–12 science education must ensure students develop the knowledge, skills, and dispositions needed to navigate and thrive in AI-supported environments.

*Core Competencies for AI in Science Education*

Educational outcomes should emphasize three interconnected competencies: (a) Knowing about AI: understanding its concepts, functions, and role in science and society, (b) Applying AI: using AI tools for inquiry, modeling, analysis, and problem-solving, and (c) Evaluating and Reflecting on AI: critically assessing AI's outputs, biases, and ethical implications.

*Reflection and Ethical Responsibility*

> Students must be equipped to move beyond surface-level tool use, cultivating habits of questioning assumptions, recognizing limitations, and addressing societal and ethical impacts embedded in AI systems.

*Learning Progressions and Pathways*

> Coherent, developmentally appropriate progressions should integrate AI outcomes into existing science frameworks (e.g., NRC Framework, NGSS), while accounting for multiple trajectories—deep technical competencies for research, applied skills for vocational fields, and reflective capacities for all citizens.

*Scientific Literacy as the Ultimate Goal*

> The aim is not to train all students as AI developers but to cultivate scientifically literate citizens who can thoughtfully and ethically engage with AI to contribute to science, work, civic life, and global challenges.

## 3. Recommendations for Future Research

Building on the insights articulated in this volume, future research must take deliberate steps to advance a human-centered vision of AI in science education. As AI continues to reshape scientific practice, pedagogy, and learning environments, research must both leverage these transformations and remain grounded in the *REP* framework introduced in this book. The following directions highlight how future work can respond to the changes AI brings while safeguarding fairness, transparency, human oversight, and respect for human values. To deepen the agenda, each recommendation is elaborated to show both the opportunities created by AI and the ethical guardrails required to ensure responsible adoption.

### 3.1. Develop Coherent Conceptual and Theoretical Frameworks for AI-Supported Science Education

Future research should refine constructs such as *Science Literacy Plus* and *AI literacy* to reflect the ways AI redefines scientific inquiry, knowledge production, and classroom learning.

Theoretical work must not only account for AI's potential to accelerate data analysis, modeling, and collaboration but also integrate REP principles such as fairness, autonomy, and meaningful use. By articulating conceptual frameworks that balance innovation with responsibility, scholars can provide educators and policymakers with clear guidance for ethically grounded practice. This work should also consider how AI disrupts traditional notions of scientific authority and democratizes knowledge, ensuring that new frameworks explicitly safeguard transparency, integrity, and the preservation of human values. Expanding this line of research could involve mapping how AI transforms epistemic practices across different scientific domains, investigating how students' conceptions of science evolve when AI becomes a partner in inquiry, and theorizing how REP values interact with curricular standards and policy mandates. Such scholarship should also link conceptual development to design-based research that tests how theoretical models play out in classrooms, allowing theories to remain dynamic and responsive to emerging challenges and opportunities.

*3.2. Advance Methodological Innovation and Build Robust Evidence in AI Contexts*

AI enables the collection of rich, multimodal data and the use of advanced analytics to study learning in unprecedented detail. Future research should harness these capabilities to design studies that explain *why* and *under what conditions* AI enhances equity, creativity, and scientific reasoning. At the same time, methodologies must embody REP values by ensuring privacy protections, transparency in analytics, and accountability in interpretation. Research that blends human-centered inquiry with AI-supported tools will be essential to building trustworthy, evidence-based insights. Methodological expansion should also include longitudinal designs that track learners across educational stages, participatory approaches that include teacher and student voices, and cross-context comparisons to reveal how AI can responsibly adapt to varied learning environments. Further growth in this area might involve the creation of cross-institutional research consortia that share anonymized datasets while adhering strictly to REP safeguards, as well as the development of open-source methodological toolkits that empower educators and researchers worldwide to study AI-supported learning responsibly. By embedding ethical protocols into every stage of methodological design, scholars can ensure that evidence-building advances both scientific understanding and human dignity.

*3.3. Prioritize Ethical, Equitable, and Human-Centered Research Agendas*

As AI tools shape assessment, personalization, and instructional decision-making, it is imperative to investigate how they influence equity and agency. Future studies should focus on reducing algorithmic bias, protecting student data, and amplifying diverse voices in classrooms. Guided by REP principles of fairness, transparency, and respect for human values, research must explore not only the risks of inequity but also strategies for designing AI systems that actively promote inclusivity. Investigating how ethical commitments are enacted in real-world settings will ensure AI adoption remains human-centered. Additional attention should be devoted to contexts outside formal schooling, such as community learning hubs and informal science spaces, where AI tools may reach vulnerable populations. By doing so, researchers can better illuminate how ethical practices scale across diverse ecosystems. Expanding this agenda requires comparative work that documents how ethics is interpreted across cultural contexts, the development of rubrics for assessing ethical dimensions of AI integration, and collaborations with ethicists and community stakeholders who can help translate REP commitments into culturally responsive practices. Such expanded efforts will position ethics not as an afterthought but as a central driver of innovation.

*3.4. Strengthen Teacher Roles and Professional Learning for Ethical AI Integration*

AI is altering classroom dynamics, shifting teachers from sole orchestrators to co-facilitators alongside students and AI systems. Research should explore how educators can critically evaluate AI tools, balance guidance with student autonomy, and retain authority over ethical decision-making. Professional learning must cultivate both technical fluency and moral judgment, aligning with REP commitments to human oversight and accountability. Studies should also examine how teachers' professional identities and agency evolve in AI-supported environments, ensuring that educators remain empowered shapers of ethical practice. Future work should additionally investigate how professional development can model ethical AI use, provide teachers with case-based simulations, and create international communities of practice that exchange lessons on balancing AI innovations with human-centered pedagogy. Expanding this research could include studying how pre-service teacher preparation programs embed AI literacy, how cross-generational mentorship models support ethical practice, and how professional learning networks can act as watchdogs ensuring adherence to REP. Such expanded work will clarify how teachers can remain both pedagogical leaders and ethical stewards in AI-supported classrooms.

*3.5. Innovate Learning Materials and Assessment Practices with Responsibility*

AI has enabled adaptive, generative, and multimodal resources that can transform science learning materials and assessments. Future research should investigate how such resources can embody authenticity, accessibility, and cultural responsiveness while aligning with REP principles of transparency, robustness, and scientific integrity. Assessments must be designed to capture higher-order competencies—such as critical evaluation of AI outputs and ethical reasoning—while resisting overreliance on automation. By prioritizing validity, fairness, and human oversight, research can ensure that AI-supported assessments foster authentic learning and reflective engagement. In addition, future agendas should examine how generative AI reshapes textbook creation, how adaptive assessments can be audited for equity, and how student reflection on AI use can itself become a learning outcome, reinforcing REP commitments through practice. Expanding this focus might involve the co-design of culturally sustaining digital resources with community partners, empirical studies testing the reliability of AI-supported grading across diverse populations, and design experiments that integrate student self-assessment of AI outputs into formative feedback cycles. Through such expansion, the field can ensure that learning resources and assessments both leverage AI's strengths and safeguard against its risks.

### *3.6. Foster Interdisciplinary and Cross-Sector Collaboration Guided by Responsible and Ethical Principles*

Because AI integration in education is embedded in larger socio-technical systems, future research should encourage partnerships across disciplines, sectors, and communities. Collaborative projects must not only advance technological innovation but also ensure alignment with REP commitments. For example, open educational resources and shared data infrastructures should be designed with transparency, accountability, and inclusivity in mind. Interdisciplinary teams—including educators, technologists, ethicists, and policymakers—can co-create solutions that are both innovative and human-centered. Research should also examine governance models for such collaborations, exploring how collective responsibility can be distributed fairly, how policies can enforce REP, and how voices from underrepresented communities can have meaningful influence in shaping the direction of AI in education. Further expansion could include studies of international governance structures for educational AI, analyses of how policy frameworks operationalize REP commitments, and experimental collaborations that test new forms of participatory design with teachers and students. By scaling up collaborative research while embedding REP safeguards, the field can chart models of AI governance that are democratic, inclusive, and future-ready.

*3.7. Embrace Global and Comparative Research Agendas to Ensure Inclusive AI Futures*

AI adoption is unfolding differently across nations and cultures, with varied implications for equity, pedagogy, and policy. Future research should compare these diverse pathways to identify practices that balance innovation with REP commitments to fairness, cultural responsiveness, and respect for human dignity. Global collaborations can generate repositories of open-source AI tools, ethical guidelines, and multilingual resources that reflect both universal principles and local adaptations. Special attention should be given to voices from the global South and marginalized communities, ensuring that the human-centered vision for AI in science education is inclusive and just. Further expansion in this area should involve joint cross-country projects, UNESCO-level policy dialogues, and knowledge exchanges that enable equitable participation of countries with fewer technological resources, thereby preventing the reproduction of global inequities. To extend this agenda, scholars might explore how indigenous knowledge systems intersect with AI tools, and how resource-constrained contexts innovate responsible low-tech AI solutions.

**4. Conclusion**

The integration of AI into science education represents not simply a technological innovation but a fundamental transformation of educational purposes, practices, and outcomes. This volume highlights how AI can enrich inquiry, personalize learning, and expand opportunities, yet its adoption must remain firmly anchored in Responsible and Ethical Principles. Guided by fairness, transparency, privacy, and respect for human values, AI should serve as a catalyst to amplify—rather than replace—human curiosity, creativity, and collaboration. We thus call for a human-centered vision of AI in science education serving learners who are scientifically literate, ethically grounded, and capable of critical engagement with AI. Teachers, as ethical stewards of practice, and policymakers, as architects of equitable systems, play essential roles in ensuring that AI enhances rather than diminishes human agency. By aligning innovation with responsibility and ethics, science education can transform the disruptive potential of AI into a force for human flourishing and civic responsibility.

**Acknowledgement**

*This material is based upon work supported by the National Science Foundation under Grant No. 2332964 (PI Zhai). Any opinions, findings, and conclusions or recommendations expressed in this material are those of the author(s) and do not necessarily reflect the views of the National Science Foundation.*

**Declaration of AI Use**

Parts of this manuscript were prepared with the assistance of generative AI tools. The AI tools were used for language editing or paragraphing of the authors' writing. The authors reviewed, revised, and verified all AI-supported content for accuracy, appropriateness, and originality. No confidential or sensitive data were entered into AI systems. The final responsibility for the content rests solely with the authors.

Zhai, X. (2024). Transforming Teachers' Roles and Agencies in the Era of Generative AI: Perceptions, Acceptance, Knowledge, and Practices. *Journal of Science Education and Technology*. https://doi.org/10.1007/s10956-024-10174-0

Zhai, X., & Krajcik, J. (2025a). Pseudo Artificial Intelligence Bias. In *Uses of Artificial Intelligence in STEM Education* (pp. 568–578). Oxford University Press. https://doi.org/10.1093/oso/9780198882077.003.0025

Zhai, X., & Krajcik, J. (2025b). *Uses of Artificial Intelligence in STEM Education*. Oxford University Press.

Zhai, X., & Nehm, R. (2023). AI and formative assessment: The train has left the station. *Journal of Research in Science Teaching, 60*(6), 1390–1398. https://doi.org/DOI: 10.1002/tea.21885
25